\documentclass[12pt,a4paper]{article}
\usepackage{indentfirst}
\usepackage{epsfig}
\usepackage{graphicx}
\usepackage[innercaption]{hvfloat}

\begin{document}
\title{Rapidity and energy dependence of the electric charge correlations in A+A collisions at the SPS energies.}

\maketitle

%\begin{center}
%  {\textbf{DRAFT 1.5}}
%\end{center}

\begin{center}
  {\textbf{The NA49 Collaboration}}
\end{center}

\vspace{0.5cm}
\noindent
C.~Alt$^{9}$, T.~Anticic$^{23}$, B.~Baatar$^{8}$,D.~Barna$^{4}$,
J.~Bartke$^{6}$, L.~Betev$^{10}$, H.~Bia{\l}\-kowska$^{20}$,
C.~Blume$^{9}$,  B.~Boimska$^{20}$, M.~Botje$^{1}$
% ...

\vspace{0.5cm}
\noindent
$^{1}$NIKHEF, Amsterdam, Netherlands. \\
% ...
$^{4}$KFKI Research Institute for Particle and Nuclear Physics, Budapest, Hungary.\\
% ...
$^{6}$Henryk Niewodniczanski Institute of Nuclear Physics, Polish Academy of Sciences, Cracow, Poland.\\
% ...
$^{8}$Joint Institute for Nuclear Research, Dubna, Russia.\\
$^{9}$Fachbereich Physik der Universit\"{a}t, Frankfurt, Germany.\\
$^{10}$CERN, Geneva, Switzerland.\\
% ...
$^{20}$Institute for Nuclear Studies, Warsaw, Poland.\\
% ...
$^{23}$Rudjer Boskovic Institute, Zagreb, Croatia.\\
% ...

\vspace{-0.5 cm}
\begin{abstract}

Results from electric charge correlations studied with the Balance Function method in A+A collisions from 20\emph{A} to 158\emph{A} GeV are presented in two different rapidity intervals: In the mid-rapidity region we observe a decrease of the width of the Balance Function distribution with increasing centrality of the collision, whereas this effect vanishes in the forward rapidity region.

Results from the energy dependence study in central Pb+Pb collisions show that the narrowing of the Balance Function expressed by the normalised width parameter \textit{W} increases with energy towards the highest SPS and RHIC energies. 

Finally we compare our experimental data points with predictions of several models. The hadronic string models UrQMD and HIJING do not reproduce the observed narrowing of the Balance Function. However, AMPT which contains a quark-parton transport phase before hadronization can reproduce the narrowing of the BF's width with centrality. This confirms the proposed sensitivity of the Balance Function analysis to the time of hadronization.

\end{abstract}

%\newpage

%%%%%%%%%%%%%%%%%%%%%%%%%%%%%%%%%%%%%
\section{Introduction}

The study of the Balance Function (BF) is motivated by the prediction that its width may be sensitive to the creation of a deconfined phase in the early stage of central nucleus-nucleus collisions \cite{Pratt,Drij,Aih}. The method is based on the observation that hadrons are produced locally in pairs of oppositely charged particle. Particles of such a pair are separated in rapidity (y) due to the initial momentum difference and secondary interactions with other particles. Particles of a pair that were created earlier are separated further in rapidity because of the expected large initial momentum difference and the long lasting rescattering phase. On the other hand, oppositely charged particles of a pair that were created later are correlated over a smaller interval $\Delta y$ in rapidity.

As in previous publications from the NA49 experiment on the BF \cite{BF_NA49}, the analysis is performed for all charged particles (predominantly pions) in order to avoid the additional complications from the identification procedure. Consequently the pseudo-rapidity ($\eta$) of charged particles is examined and the BF is defined as \cite{Pratt}:

\begin{center}
\begin{equation}
B(\Delta \eta) = \frac{1}{2} \Big[ \frac{N_{+-}(\Delta \eta) -
N_{--}(\Delta \eta)}{N_{-}} + \frac{N_{-+}(\Delta \eta) -
N_{++}(\Delta \eta)}{N_{+}}  \Big].
\label{BF_DEF2}
\end{equation}
\end{center}

\vspace{0.2 cm}

The width of the BF can be characterized by the weighted average $\langle \Delta \eta \rangle$:

\begin{center}
\begin{equation}
\langle \Delta \eta \rangle = \sum_{i=0}^k{(B_i \cdot \Delta \eta _i)}/\sum_{i=0}^k{B_i},
\label{width}
\end{equation}
\end{center}

\noindent where \emph{i} is the bin number of the BF histogram.

It was suggested that the creation of a deconfined phase of quarks and gluons (the QGP) in the early stages of nucleus-nucleus collisions, would lead to delayed hadronization, i.e. a narrowing of the BF \cite{Pratt}. Indeed, results on the BF obtained for Au+Au collisions at RHIC \cite{STAR} and Pb+Pb interactions at the top SPS energy \cite{BF_NA49} observed this narrowing for central collisions in the mid-rapidity domain. In this paper, the study of the BF is extended to the forward rapidity region. In addition, the energy dependence of the width of the BF for central Pb+Pb collisions is investigated at lower SPS energies where indications for the onset of deconfinement had been obtained \cite{Onset}.

In the following two sections we will first describe in brief the NA49 experimental setup and the data analysis. The next two sections will be dedicated to the new experimental results on the rapidity and the energy dependence of the width of the BF. We will conclude with a section on model comparisons and the summary.

%%%%%%%%%%%%%%%%%%%%%%%%%%%%%%%%%%%%%
\section{Experimental Setup}

The NA49 detector \cite{na49_nim} is a wide acceptance hadron spectrometer for the study of hadron production in collisions of hadrons or heavy ions at the CERN SPS. The main components are four large - volume Time Projection Chambers (TPCs) which are capable of detecting 80\% of some 1500 charged particles created in a central Pb+Pb collision at 158\emph{A} GeV. 

The targets are C (561 mg/cm$^{2}$), Si (1170 mg/cm$^{2}$) disks and a Pb (224 mg/cm$^{2}$) foil for ion collisions and a liquid hydrogen cylinder (length 20 cm) for hadron interactions. They are positioned about 80 cm upstream from VTPC-1.

The centrality of a collision is selected (on-line for central Pb+Pb, Si+Si and C+C and off-line for minimum bias Pb+Pb, Si+Si and C+C interactions) by a trigger using information from a downstream calorimeter (VCAL), which measures the energy $E_0$ of the projectile spectator nucleons.

%%%%%%%%%%%%%%%%%%%%%%%%%%%%%%%%%%%%%%%%%%%%
\section{Data Analysis}

%%%%%%%%%%%%%%%%%%%%%%%%%%%%%%%%%%%%%%%%%%%%
\subsection{Data Sets}

\begin{table}[ht]
\begin{center}
\begin{tabular}{|c|c|c|c|c|c|}
\hline  Interaction & Number of events & $\sigma_{cent.} / \sigma_{inel.} [\%]$& $\langle N_{W} \rangle$ \\
\hline 
\hline p+p      & 200K &             &     \\
\hline C+C      & 45K  & 15.3        & 14  \\
\hline Si+Si    & 65K  & 12.2        & 37  \\
\hline Pb+Pb(6) & 38K  & 43.5 - ...  & $45 \pm 4$  \\ 
\hline Pb+Pb(5) & 20K  & 33.5 - 43.5 & $85 \pm 7$ \\ 
\hline Pb+Pb(4) & 24K  & 23.5 - 33.5 & $128 \pm 8$ \\ 
\hline Pb+Pb(3) & 24K  & 12.5 - 23.5 & $196 \pm 6$ \\ 
\hline Pb+Pb(2) & 50K  & 5 - 12.5    & $281 \pm 4$ \\ 
\hline Pb+Pb(1) & 38K  & 0 - 5       & $352 \pm 3$ \\

\hline
\end{tabular}
\end{center}
\caption{Systems and centrality classes used in this analysis. Listed for p+p, C+C, Si+Si and six centralities of Pb+Pb collisions at 158\emph{A} GeV are the number of events, the cross section of selected central interactions as percentage of the total inelastic cross section and the mean number $\langle N_{W} \rangle$ of wounded nucleons.}
\label{DataSetsSystem160}
\end{table}

The data sets used in the rapidity dependence study come from p+p, C+C, Si+Si and Pb+Pb collisions at 158\emph{A} GeV (Table \ref{DataSetsSystem160}) and at 40\emph{A} GeV (Table \ref{DataSetsSystem40}). For Pb+Pb interactions, data with both central and minimum bias trigger have been analyzed in order to study the centrality dependence of the BF. The minimum bias data were subdivided into six different centrality classes \cite{Cooper} according to the energy recorded by the VCAL, from class Veto 1 (the most central collisions) to class Veto 6 (the most peripheral collisions) (Table \ref{DataSetsSystem160} and \ref{DataSetsSystem40}). For the energy dependence study, we have analyzed central Pb+Pb collisions throughout the whole SPS energy range (Table \ref{DataSetsEnergy}). The most central Pb+Pb interactions for the highest SPS energy correspond to $5\%$ whereas for lower energies to $7.2\%$ of the total geometric cross section. 

\begin{table}[ht]
\begin{center}
\begin{tabular}{|c|c|c|c|c|c|}
\hline  Interaction & Number of events & $\sigma_{cent.} / \sigma_{inel.} [\%]$& $\langle N_{W} \rangle$ \\
\hline 
\hline C+C      & 76K  & 65.7        & 9   \\
\hline Si+Si    & 43K  & 29.2        & 32  \\
\hline Pb+Pb(6) & 45K  & 43.5 -      & $43 \pm 4$  \\ 
\hline Pb+Pb(5) & 27K  & 33.5 - 43.5 & $93 \pm 7$ \\ 
\hline Pb+Pb(4) & 30K  & 23.5 - 33.5 & $142 \pm 8$ \\ 
\hline Pb+Pb(3) & 32K  & 12.5 - 23.5 & $210 \pm 6$ \\ 
\hline Pb+Pb(2) & 106K & 5 - 12.5    & $290 \pm 4$ \\ 
\hline Pb+Pb(1) & 190K & 0 - 5       & $351 \pm 3$ \\

\hline
\end{tabular}
\end{center}
\caption{Systems and centrality classes used in this analysis. Listed for C+C, Si+Si and six centralities of Pb+Pb collisions at 40\emph{A} GeV are the number of events, the cross section of selected central interactions as percentage of the total inelastic cross section and the mean number $\langle N_{W} \rangle$ of wounded nucleons.}
\label{DataSetsSystem40}
\end{table}

\begin{table}[ht]
\begin{center}
\begin{tabular}{|c|c|c|c|c|}
\hline  $\sqrt{s_{NN}}$ [GeV] &  $E_{beam}$ [\textit{A} GeV] & $\sigma_{cent.} / \sigma_{inel.} [\%]$ & $\langle N_{W} \rangle$ \\
\hline 
\hline 6.3  &  20 & 7 & $349 \pm 1 \pm 5$ \\
\hline 7.6  &  30 & 7 & $349 \pm 1 \pm 5$ \\
\hline 8.8  &  40 & 5 & $361 \pm 1 \pm 5$ \\
\hline 12.3 &  80 & 7 & $349 \pm 1 \pm 5$ \\
\hline 17.3 & 158 & 5 & $362 \pm 8$ \\

\hline
\end{tabular}
\end{center}
\caption{The different energies for Pb+Pb interactions used to perform the energy scan. The cms energy of the system as well as the different beam energies, the cross section of selected central interactions as percentage of the total inelastic cross section as well as the number of wounded nucleons are listed. The third number in the last column refers to the statistical error whereas the second to the systematic.}
\label{DataSetsEnergy}
\end{table}

%%%%%%%%%%%%%%%%%%%%%%%%%%%%%%%%%%%%%%
\subsection{Event and Track Selection}

In order to reduce the contamination from non-target events and non-vertex tracks, selection criteria were imposed both at the event and the track level.

Events were selected that had a proper position of the reconstructed primary vertex. The vertex coordinate $V_z$ along the beam axis had to fulfill $|V_z - V_{z_0}| < \Delta z$ where the values of the central position $V_{z_0}$ and the range $\Delta z$ vary with respect to the system and the energy analyzed. In addition the vertex coordinates $V_x$ and $V_y$ perpendicular to the beam axis had to fulfill $|V_x - V_{x_0}| < \Delta x$ and $|V_y - V_{y_0}| < \Delta y$, where the values $V_{x_0}, V_{y_0}$ and $\Delta x$, $\Delta y$ also vary accordingly.

Selection criteria at the track level were imposed in order to reduce the contamination by tracks from weak decays, secondary interactions and other sources of non-vertex tracks. Thus, an accepted track had to have an extrapolated distance of closest approach $d_x$ and $d_y$ of the particle at the vertex plane within the range $|d_x| < 2.0$ cm and $|d_y| < 1.0$ cm. In addition the potential number of points in the detector for the selected tracks had to be more than 30. To suppress double counting due to track splitting the ratio of the number of reconstructed points to the potential number of points was required to be larger than $0.5$.

Finally, the NA49 detectors provide large acceptance in momentum space; however the acceptance in the azimuthal angle $\phi$ is not complete. To avoid edge effects and to facilitate comparisons with model calculations, particles were selected in a restricted acceptance region \cite{Jacek}. The boundary of this region is described by Eq. \ref{Acc_Jac}:

\begin{equation}
p_{T}(\phi) = \frac{1}{A+(\frac{D+\phi}{C})^{6}}+B,
\label{Acc_Jac}
\end{equation}

\noindent where the values of the parameters $A$, $B$, $C$ and $D$ depend on the rapidity interval and on the energy and can be found in \cite{Jacek}.

%%%%%%%%%%%%%%%%%%%%%%%%%%%%%%%%%%%%%
\section{Rapidity Dependence}

\begin{figure}
\includegraphics[height=.3\textheight,width=0.7\textheight]{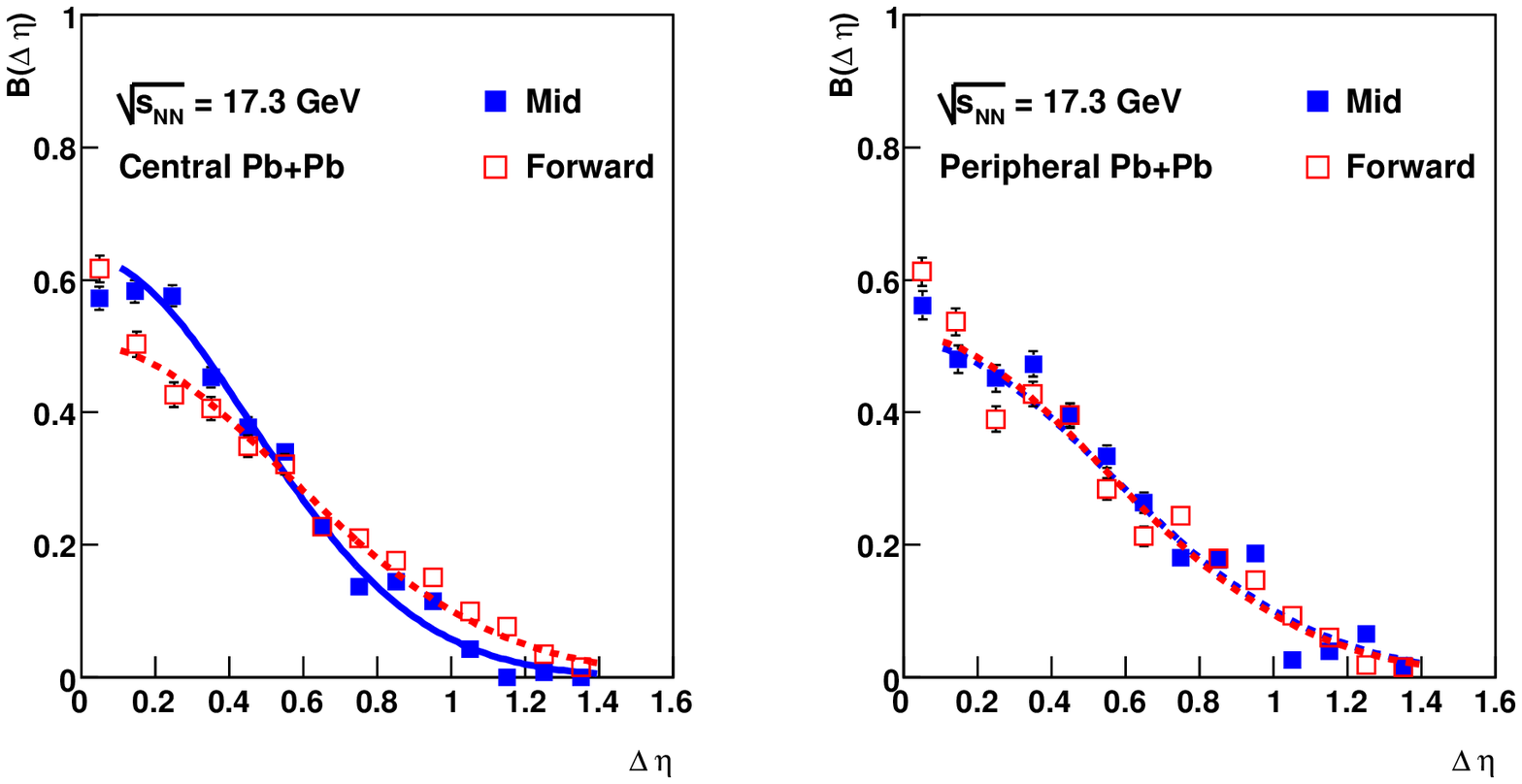}
\includegraphics[height=.3\textheight,width=0.7\textheight]{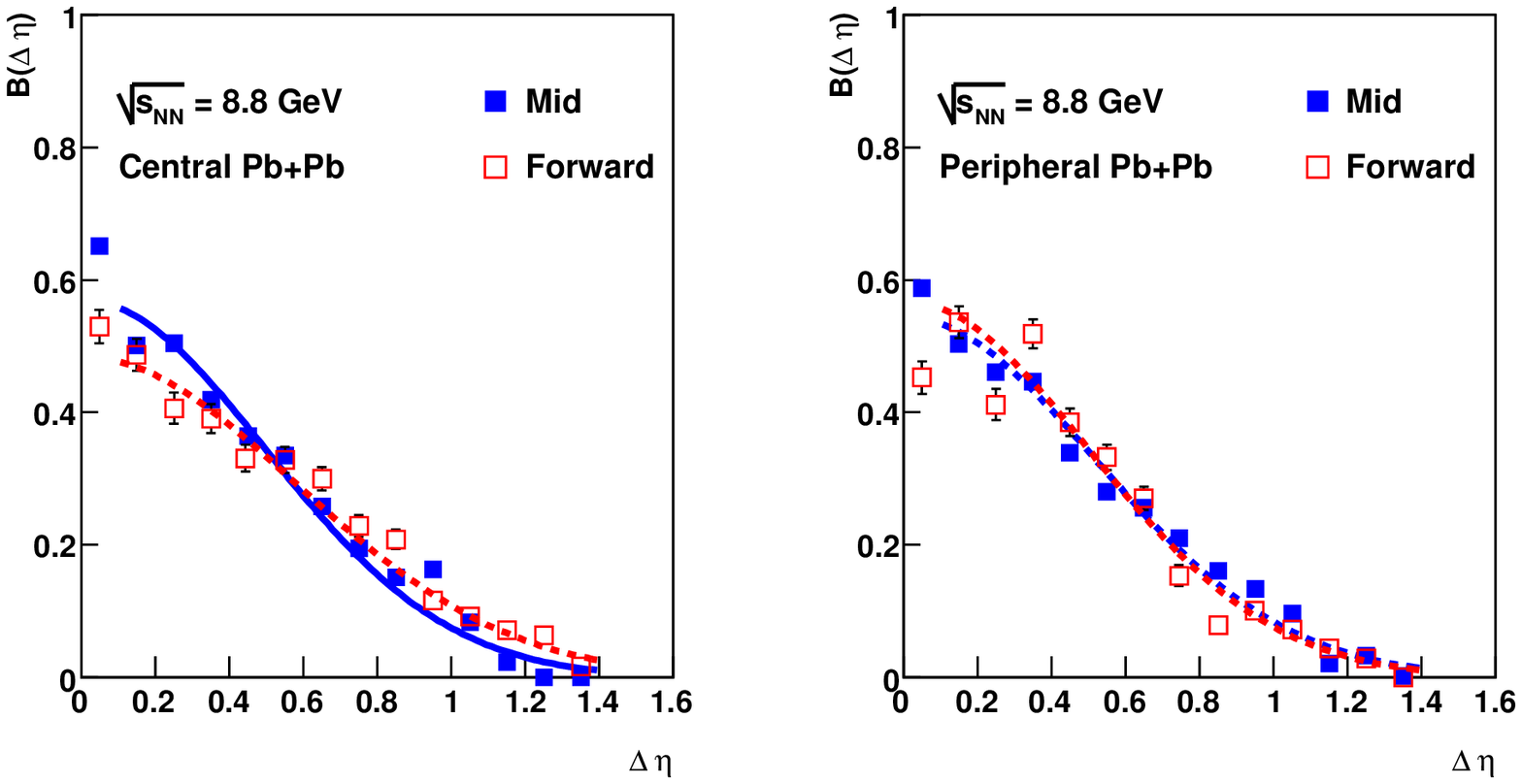}
\caption{The BF as a function of the pseudo-rapidity difference $\Delta \eta$ for central (centrality class 1 of Tables \ref{DataSetsSystem160} and \ref{DataSetsSystem40} - left plots) and peripheral (centrality class 6 of Tables \ref{DataSetsSystem160} and \ref{DataSetsSystem40} - right plots) Pb+Pb collisions at $\sqrt{s_{NN}} = 17.3$ GeV (upper plots) and $\sqrt{s_{NN}} = 8.8$ GeV (lower plots) for the two rapidity regions. The curves drawn are Gaussian fits used to guide the eye.}
\label{rapidity}
\end{figure}

\begin{figure}
\begin{center}
\includegraphics[height=.3\textheight,width=0.3\textheight]{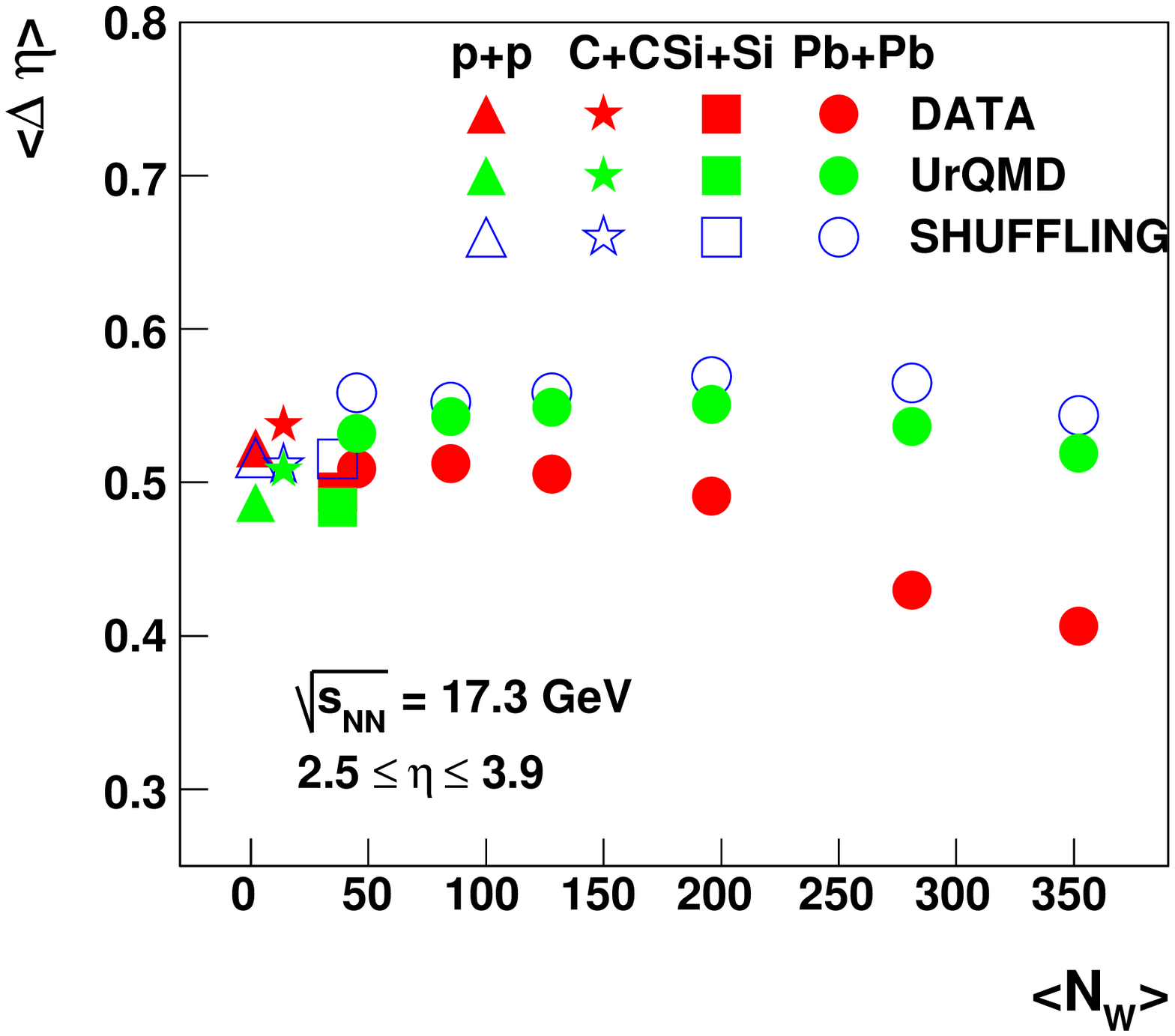}
\includegraphics[height=.3\textheight,width=0.3\textheight]{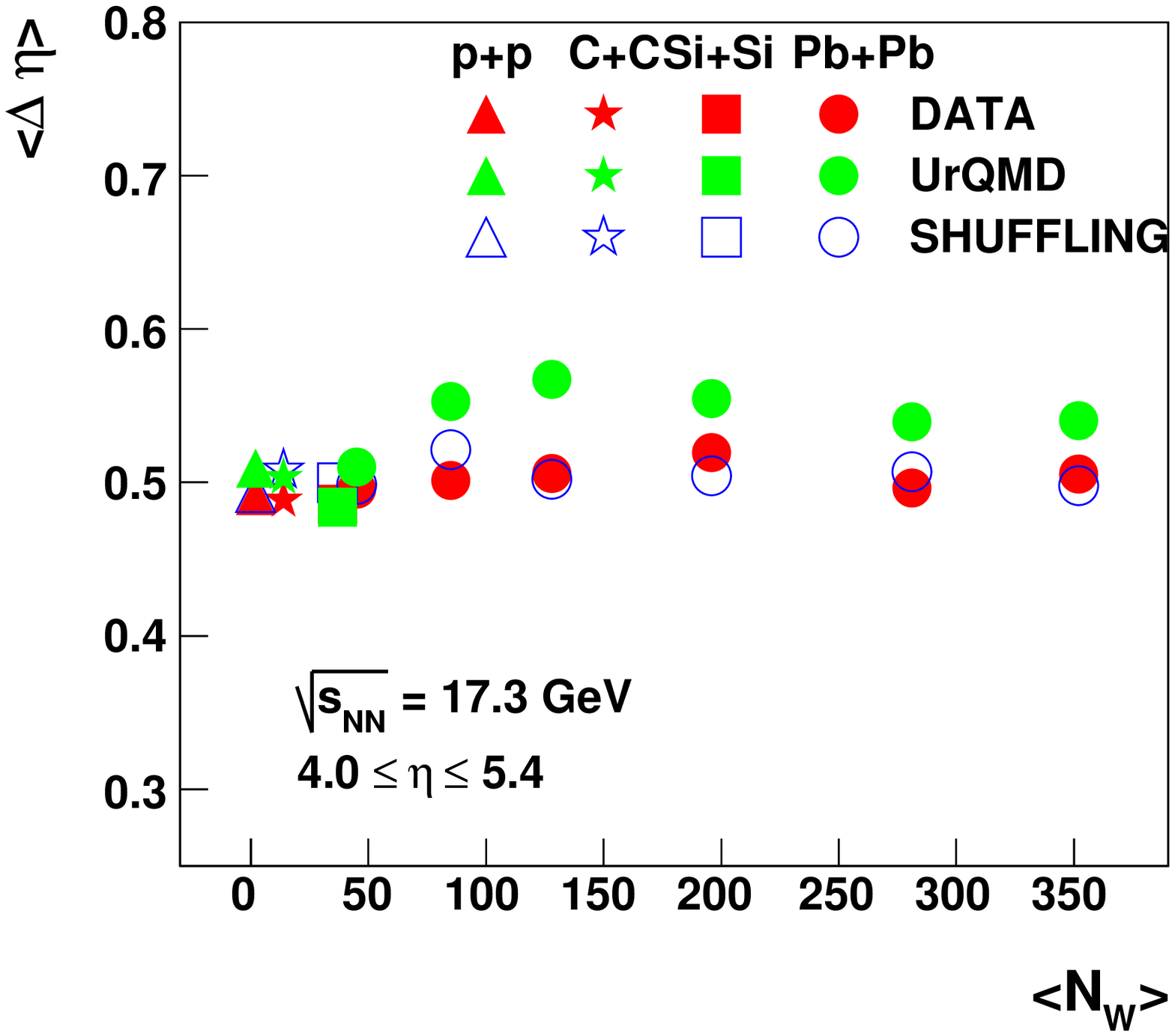}
\includegraphics[height=.3\textheight,width=0.3\textheight]{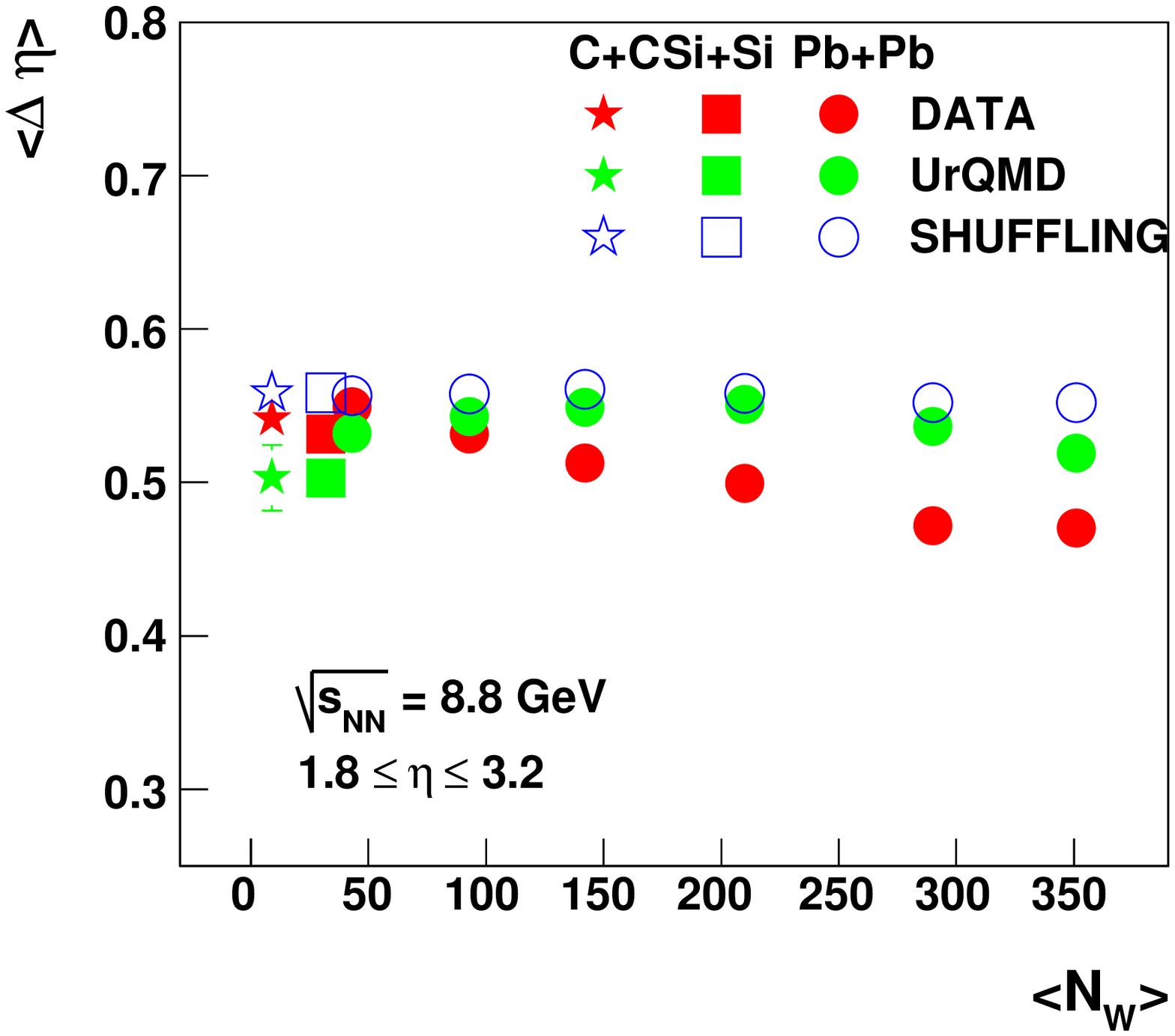}
\includegraphics[height=.3\textheight,width=0.3\textheight]{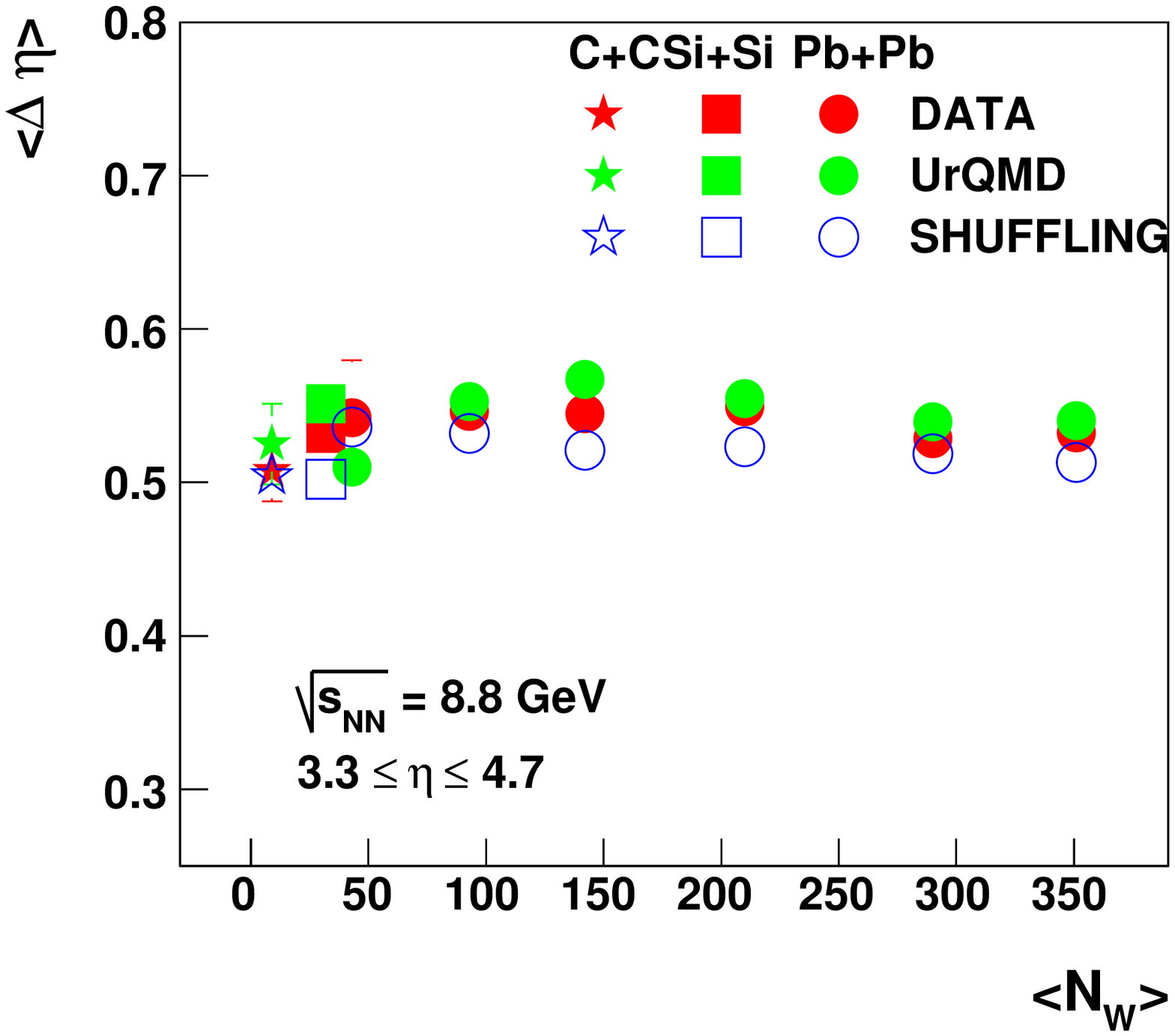}
\caption{The system size and centrality dependence of the measured width $\langle \Delta \eta \rangle$ of the Balance Function for charged particles at $\sqrt{s_{NN}} = 17.3$ GeV (top plots) and $\sqrt{s_{NN}} = 8.8$ GeV (bottom plots) plotted as a function of the mean number of wounded nucleons $\langle N_{w} \rangle$ for two different rapidity regions (Table \ref{intervals}): the mid-rapidity region (left plots) and the forward rapidity region (right plots). }
\label{NA49_rap_160}
\end{center}
\end{figure}

In order to investigate the properties of hadronization in heavy ion collisions, as proposed by the BF methodology \cite{Pratt}, we have studied the system size and centrality dependence of the width of the BF at two SPS energies ($158A$ and $40A$ GeV, corresponding to $\sqrt{s_{NN}} = 17.3$ and $8.8$ GeV, respectively) and in two different pseudo-rapidity intervals. The selected pseudo-rapidity regions for $\sqrt{s} = 17.3$ GeV are $2.5 \leq \eta \leq 3.9$ named as mid-rapidity region and $4.0 \leq \eta \leq 5.4$ named as forward rapidity region. The corresponding pseudo-rapidity regions for $\sqrt{s} = 8.8$ GeV are $1.8 \leq \eta \leq 3.2$ (mid-rapidity region) and $3.3 \leq \eta \leq 4.7$ (forward rapidity region). Fig. \ref{rapidity} shows the resulting BF distributions for the two different energies (upper and lower plots respectively) as a function of the pseudo-rapidity difference $\Delta \eta$ for central (left plots) and peripheral (right plots) Pb+Pb collisions.

The top plots of fig. \ref{NA49_rap_160} shows the width of the BF distributions for experimental data as a function of the mean number of wounded nucleons, for the two pseudo-rapidity regions (mid-rapidity region-left plot, forward rapidity region-right plot) for the $\sqrt{s_{NN}} = 17.3$ GeV case. The figure indicates that there is a clear centrality dependence for real data in the mid-rapidity region while this dependence vanishes in the forward region. In order to further investigate this effect, the Ultra-relativistic Quantum Molecular Dynamics model (UrQMD) \cite{Urqmd}, which is a microscopic model of (ultra)relativistic heavy ion collisions in the energy range from Bevalac and SIS up to AGS, SPS and RHIC, was used. The results from the study of this conventional hadronic model as well as the widths obtained from shuffled\footnote{In shuffled events the particle (pseudo)rapidities in the event are randomly reassigned while the particle charges are unchanged. This procedure destroys (pseudo)rapidity correlations but retains global charge conservation effects.} data \cite{STAR,BF_NA49} are also plotted in fig. \ref{NA49_rap_160}. No system size and centrality dependence, regardless of the rapidity region, is seen for either UrQMD or shuffled events. The same behavior is observed for $\sqrt{s_{NN}} = 8.8$ GeV in the bottom plots of fig. \ref{NA49_rap_160}. 

One should also note that there is no significant difference in the actual values of the widths between the two rapidity regions at both energies for the small systems, such as p+p up to Si+Si. In other words, the difference in the absolute values appears only for the most central Pb+Pb collisions at both energies.

%%%%%%%%%%%%%%%%%%%%%%%%%%%%%%%%%%%%%%%%%%%%
\section{Energy Dependence}

\begin{table}[ht]
\begin{center}
\begin{tabular}{|c|c|c|}
\hline  $\sqrt{s_{NN}} [GeV]$ &  $\eta$ interval (mid-rapidity) &  $\eta$ (forward rapidity) \\
\hline 
\hline 6.3 & $1.6 \leq \eta \leq 3.0$ & $3.1 \leq \eta \leq 4.5$\\
\hline 7.6 & $1.7 \leq \eta \leq 3.1$ & $3.2 \leq \eta \leq 4.6$\\
\hline 8.8 & $1.8 \leq \eta \leq 3.2$ & $3.3 \leq \eta \leq 4.7$\\
\hline 12.3 & $2.2 \leq \eta \leq 3.6$ & $3.7 \leq \eta \leq 5.1$\\
\hline 17.3 & $2.5 \leq \eta \leq 3.9$ & $4.0 \leq \eta \leq 5.4$\\

\hline
\end{tabular}
\end{center}
\caption{The table lists the definition of the mid-rapidity as well as the forward rapidity interval for the different SPS energies.}
\label{intervals}
\end{table}

The energy dependence of the BF was studied for central Pb+Pb collisions for which NA49 data is available throughout the SPS energy range (100K events were analyzed for each energy). These data samples passed once again through the shuffling mechanism to provide an estimate of the highest value of the width for each energy under the constraint of global charge conservation. The pseudo-rapidity interval analyzed was limited to a range of 1.4 units and was located around mid-rapidity (Table \ref{intervals}). In order to compare the decrease of the width for the different energies, we introduced the normalized parameter \textit{W} which is defined by the following equation:

\begin{center}
\begin{equation}
W = \frac{100 \cdot (\langle \Delta \eta \rangle_{shuffled} - \langle \Delta \eta \rangle_{data})}{\langle \Delta \eta \rangle_{shuffled}}.
\label{W}
\end{equation}
\end{center}

\begin{figure}
\begin{center}
\includegraphics[height=.3\textheight,width=1.2\textwidth]{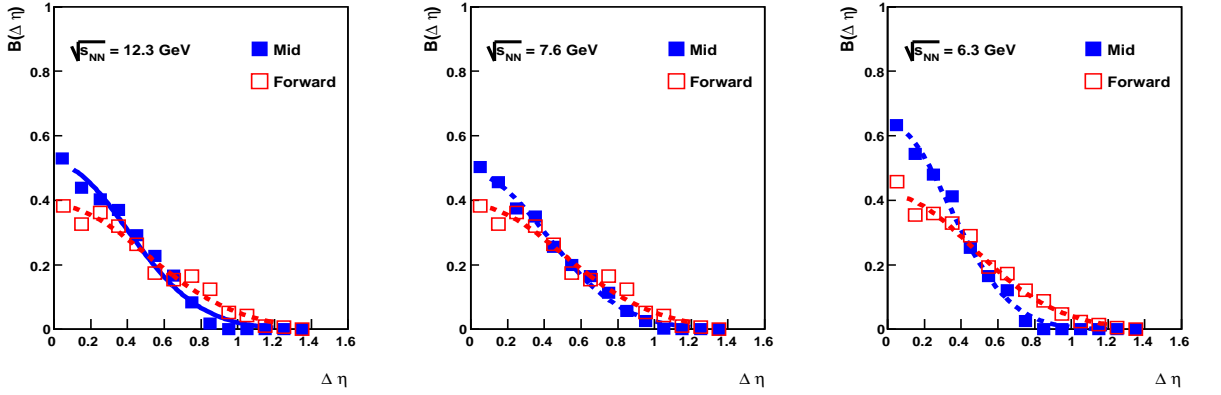}
\caption{The BF as a function of the pseudo-rapidity difference $\Delta \eta$ for central Pb+Pb collisions at $\sqrt{s_{NN}} = 12.3$ GeV (left plot), $\sqrt{s_{NN}} = 7.6$ GeV (middle plot) and $\sqrt{s_{NN}} = 6.3$ GeV (right plot) for the two rapidity regions. The curves drawn are Gaussian fits used to guide the eye.}
\label{rapidity_energy}
\end{center}
\end{figure}

By using the measure \textit{W}, we try to quantify the change of the BF width due to effects other than those of global charge conservation. Moreover, the slight changes of acceptance with energy should approximately cancel. Fig. \ref{rapidity_energy} shows as filled squares the BF distributions as a function of the pseudo-rapidity difference $\Delta \eta$ for central Pb+Pb collisions at $\sqrt{s_{NN}} = 12.3$ GeV (left plot), $\sqrt{s_{NN}} = 7.6$ GeV (middle plot) and $\sqrt{s_{NN}} = 6.3$ GeV (right plot). Fig. \ref{rapidity} shows the results for the other two energies $\sqrt{s_{NN}} = 17.3$ GeV and $\sqrt{s_{NN}} = 8.8$ GeV.

\begin{figure}
\includegraphics[height=.3\textheight]{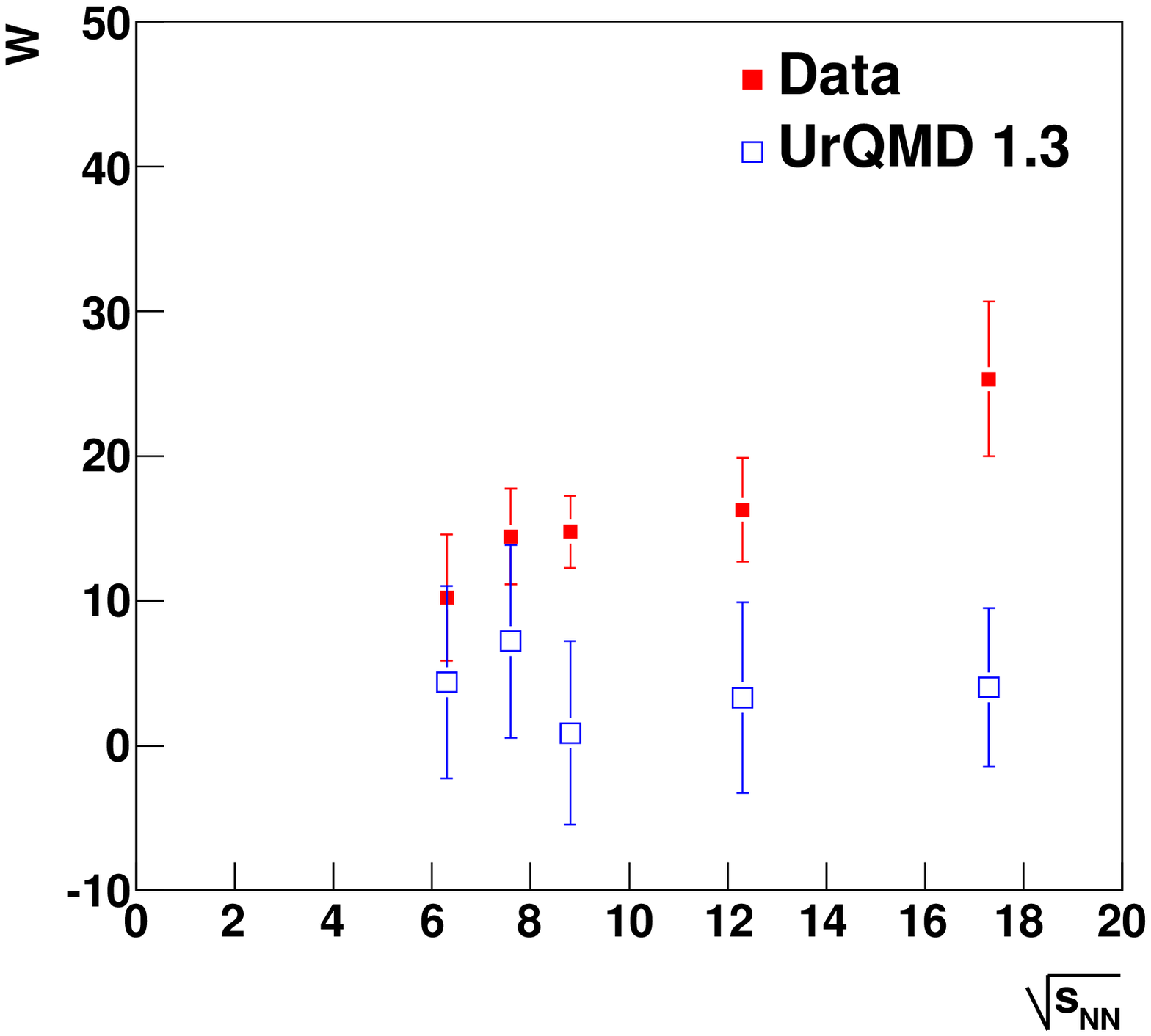}
\includegraphics[height=.3\textheight]{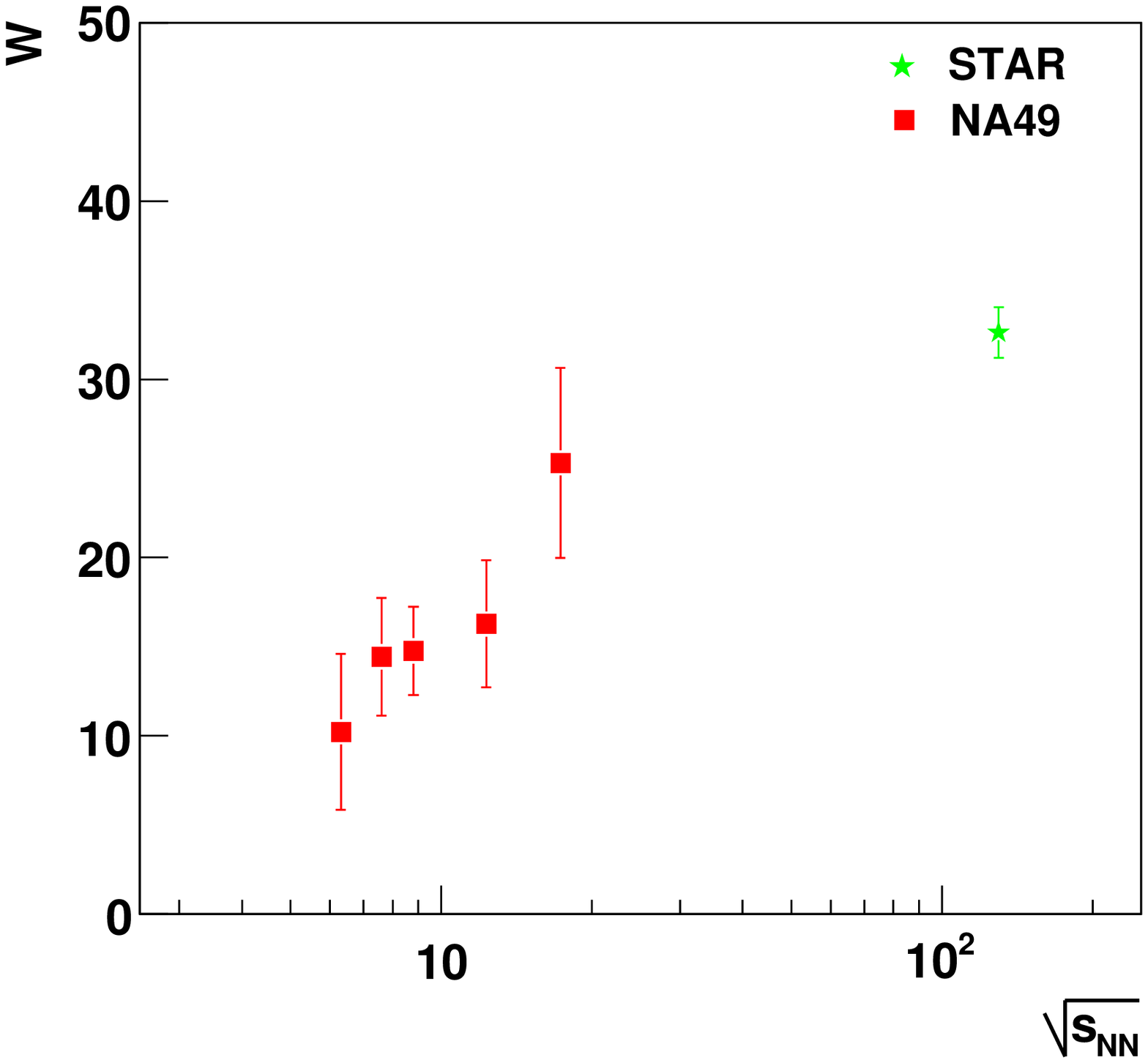}
\caption{The dependence of the normalized parameter \textit{W} on the $\sqrt{s_{NN}}$ for central Pb+Pb collisions in the SPS energy range after applying the NA49 acceptance filter (left plot) and for central Au+Au collisions at RHIC (right plot).}
\label{energy_NA49}
\end{figure}

The left plots of fig. \ref{energy_NA49} shows the dependence of the normalized \textit{W} parameter on $\sqrt{s_{NN}}$. For the data (red squares in Fig. \ref{energy_NA49}), we note an indication for an increase with energy. In contrast, the results from the analysis of UrQMD generated events (blue boxes in Fig. \ref{energy_NA49}), show no significant energy dependence of the measure \textit{W}.

\hvFloat[floatPos=htb,capWidth=0.5,capPos=r]{figure}{\includegraphics[height=.3\textheight]{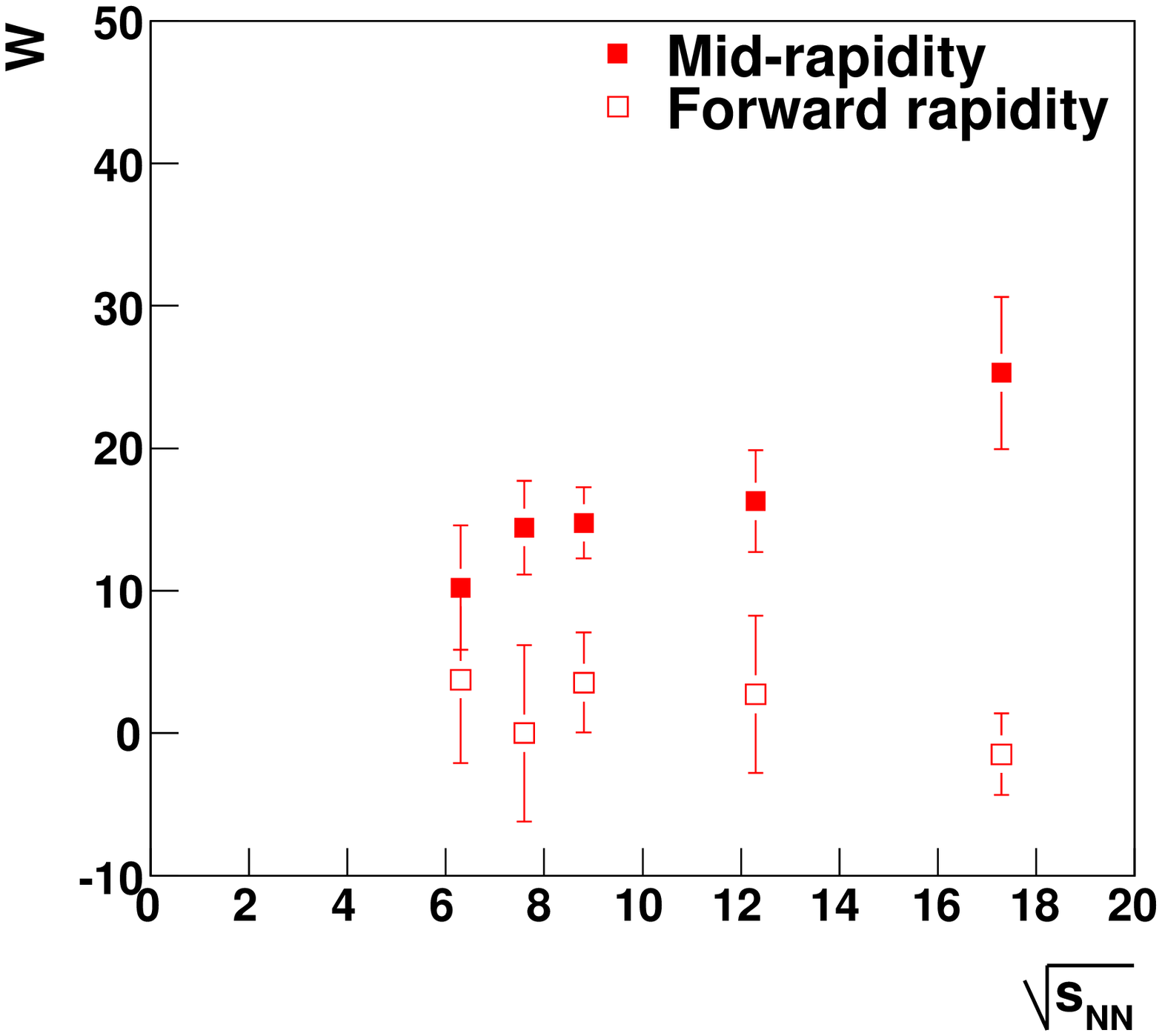}}{The dependence of the normalized parameter \textit{W} on the $\sqrt{s_{NN}}$ for central Pb+Pb collisions in the SPS energy range after applying the NA49 acceptance filter for the mid-rapidity and for the forward rapidity region (the intervals analyzed are given in Table \ref{intervals}).}{energy_NA49_forward}

The right plot of fig. \ref{energy_NA49} shows the dependence of the normalized \textit{W} parameter on $\sqrt{s_{NN}}$ for the energy range of SPS together with the RHIC result at $\sqrt{s_{NN}} = 130$ GeV \cite{STAR}. The corresponding RHIC point tends to be even higher than the last SPS one, indicating a continued rise of the \textit{W} parameter towards RHIC leaving an open question for the LHC energies.

Finally, we study the energy dependence of the scaled parameter \textit{W} of the BF's width in the forward rapidity region (Table \ref{intervals}). The corresponding BF distributions for central Pb+Pb collisions are shown as open squares in Fig. \ref{rapidity_energy} at $\sqrt{s_{NN}} = 12.3$ GeV (left plot), $\sqrt{s_{NN}} = 7.6$ GeV (middle plot), $\sqrt{s_{NN}} = 6.3$ GeV (right plot) and in Fig. \ref{rapidity} at $\sqrt{s_{NN}} = 8.8$ GeV (bottom left) and $\sqrt{s_{NN}} = 17.3 GeV$ (top left). Fig. \ref{energy_NA49_forward} summarizes the main result from this study: The normalized parameter \textit{W} shows no sign of an energy dependence in the forward rapidity regions.

%%%%%%%%%%%%%%%%%%%%%%%%%%%%%%%%%%%%%
\section{Model comparison}

In an attempt to interpret the results obtained for the centrality dependence of the BF width, several microscopic models were analyzed such as HIJING \cite{Hijing}, UrQMD \cite{Urqmd} and AMPT \cite{Ampt}. The first is based on the excitation of strings and their subsequent hadronization according to the LUND model \cite{Lund}. The second, is also based on string excitation with the addition of interactions between the produced particles after the hadronization (rescattering). The last is a multi-phase transport model which contains a quark-parton transport phase before hadronization. In AMPT, the initial spatial and momentum distributions of partons and string excitations are obtained from HIJING. The parton cascade follows the ZPC model \cite{Zpc}. When the interactions of the partons cease, they are recombined with their parent strings to form hadrons according to the LUND fragmentation mechanism \cite{Lund}. Then the scatterings of the produced hadrons are described by the ART model \cite{Art}.

Figure \ref{model_comparison}, in which the measured width $\langle \Delta \eta \rangle$ of the BF for charged particles at $\sqrt{s_{NN}} = 17.3$ GeV is plotted as a function of the mean number of wounded nucleons $\langle N_{w} \rangle$, shows the main results of our comparison. The clear centrality dependence of the order of $17 \pm 3\%$ seen for experimental data, cannot be described by the microscopic models HIJING and UrQMD. Results from both models suggest no centrality dependence of the BF width. On the other hand, the AMPT model, which incorporates the time evolution of partons before their hadronization, is predicting a centrality dependence of the BF similar to what is seen in the data. %Additional studies were performed to investigate the possible multiplicity dependence of the narrowing of the width for the AMPT model, which revealed no such dependence.

\hvFloat[floatPos=htb,capWidth=0.5,capPos=r]{figure}{\includegraphics[height=.3\textheight]{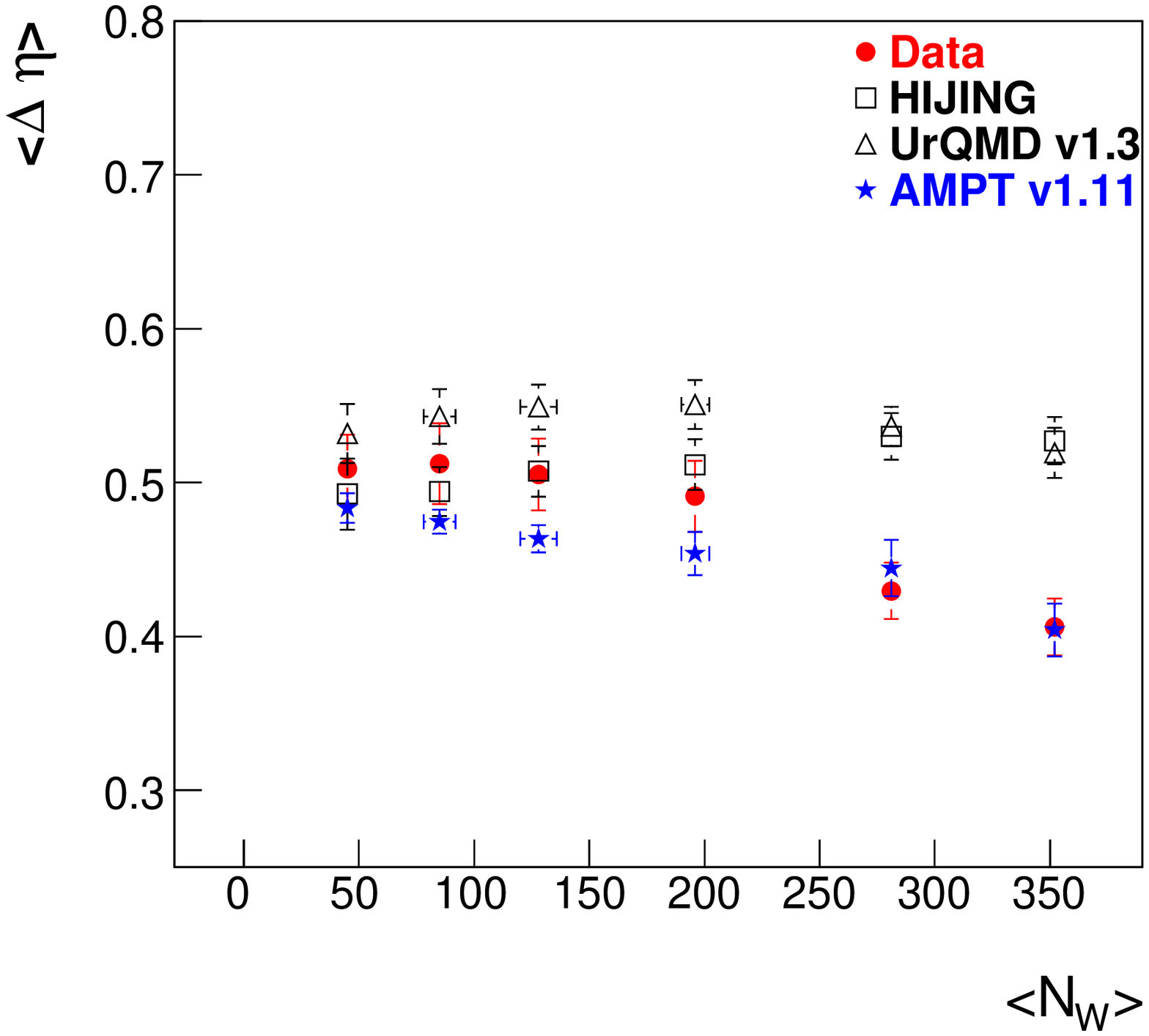}}{The centrality dependence of the measured width $\langle \Delta \eta \rangle$ of the Balance Function for charged particles at $\sqrt{s_{NN}} = 17.3$ GeV as a function of the mean number of wounded nucleons $\langle N_{w} \rangle$. The plot shows the experimental data along with the data points from different microscopic models. The interval analyzed can be seen in Table \ref{intervals}.}{model_comparison}

More comprehensive model comparisons are in progress and will be addressed in a future publication.

%%%%%%%%%%%%%%%%%%%%%%%%%%%%%%%%%%%%%
\section{Summary}

In this paper, we have presented new results on the BF obtained by the NA49 collaboration. We have analyzed data from different colliding systems and centrality classes for two different energies in two different rapidity regions. At mid-rapidity the experimental data reveal an interesting decrease of the BF width with increasing centrality of the collision for both energies. This effect can not be reproduced either by the UrQMD model or by the shuffling mechanism. In the forward rapidity region, on the other hand, no centrality dependence of the BF width is observed.

The results from the energy scan show an indication for an energy dependence which is not apparent in the UrQMD model. The narrowing of the Balance Function, expressed by the normalised width parameter \textit{W}, increases from the lowest SPS energy to the highest SPS and RHIC energies. Again this rise is limited to the mid-rapidity regions.

Finally, the comparison with different microscopic models, showed that the narrowing of the BF's width with centrality can only be reproduced by models that contain a quark-parton evolution phase before hadronization. The latter might be a first indication that indeed the BF is related to the time of hadronization.

%The centrality and energy dependence of the charge correlations of identified particles as well as an attempt to interpret the effects by comparing the experimental results with thermal models will be addressed in a future publication.

\vspace{0.5 cm}
\noindent \textbf{This paper is dedicated to the memory of Angelos Petridis.}
\vspace{0.5 cm}

%%%%%%%%%%%%%%%%%%%%%%%%%%%%%%%%%%%%%
{\bf{Acknowledgments}}

This work was supported by the University of Athens project PYTHAGORAS II - Support of Univ. Research Groups, the US Department of Energy Grant DE-FG03-97ER41020/A000, the Bundesministerium fur Bildung und Forschung, Germany, the Virtual Institute VI-146 of Helmholtz Gemeinschaft, Germany, the Polish State Committee for Scientific Research (1 P03B 006 30, 1 P03B 097 29, 1 PO3B 121 29, 1 P03B 127 30), the Hungarian Scientific Research Foundation (T032648, T032293, T043514), the Hungarian National Science Foundation, OTKA, (F034707), the Polish-German Foundation, the Korea Science \& Engineering Foundation (R01-2005-000-10334-0), the Bulgarian National Science Fund (Ph-09/05) and the Croatian Ministry of Science, Education and Sport (Project 098-0982887-2878).

\end{document}